\documentclass[intlimits,twoside,a4paper]{article}

\usepackage{lmodern}

\usepackage{amsfonts}
\usepackage{graphicx}
\usepackage{epsfig}

\usepackage{amsmath}
\usepackage{amssymb}
\usepackage{amsbsy}
\usepackage{esint}
\usepackage{tikz} 
\usetikzlibrary{arrows}
\usepackage{latexsym}
\usepackage{appendix}
\usepackage{makeidx}
\usepackage{url}
\usepackage{fancyhdr}
\usepackage{subfigure}
\usepackage{float} 
\usepackage{setspace}
\usepackage{mathrsfs}

\usepackage[T2A]{fontenc}
\usepackage[cp1251]{inputenc}

\usepackage[eqsecnum]{cmpj2}

\issue{2015}{18}{2}{23002}
\doinumber{10.5488/CMP.18.23002}

\title[Meron-Cluster simulation of the quantum Heisenberg model]%
{Meron-cluster simulation of the quantum antiferromagnetic Heisenberg model
in a magnetic field in one- and two-dimensions}
\author[G. Palma, A. Riveros]{G. Palma,  A. Riveros}
\address{
Departamento de F\'{\i}sica, Universidad de Santiago de Chile,
Casilla 307, Santiago 2, Chile
}

\date{Received October 9, 2014, in final form March 4, 2015}
\authorcopyright{G. Palma, A. Riveros, 2015}

\begin{document}

\maketitle

\begin{abstract}
Motivated by the numerical simulation of systems which display quantum phase transitions, we present a novel application of the meron-cluster algorithm to simulate the quantum antiferromagnetic Heisenberg model coupled to an external uniform magnetic field both in one and in two dimensions. In the infinite volume limit and at zero temperature we found numerical evidence that supports a quantum phase transition very close to the critical values $B_\textrm{c}=2$ and $B_\textrm{c} = 4$ for the system in one and two dimensions, respectively. For the one dimensional system, we have compared the numerical data obtained with analytical predictions for the magnetization density as a function of the external field obtained by scaling-behaviour analysis and Bethe Ansatz techniques. Since there is no analytical solution for the two dimensional case, we have compared our results with the magnetization density obtained by scaling relations for small lattice sizes and with the approximated thermodynamical limit at zero temperature guessed by scaling relations. Moreover, we have compared the numerical data with other numerical simulations performed by using different algorithms in one and two dimensions, like the directed loop method. The numerical data obtained are in perfect agreement with all these previous results, which confirms that the meron-algorithm is reliable for quantum Monte Carlo simulations and applicable both in one and two dimensions. Finally, we have computed the integrated autocorrelation time to measure the efficiency of the meron algorithm in one dimension.
\pacs 05.50.+q, 02.70.-c, 68.35.Rh, 75.40.Cx
\end{abstract}

\section{Introduction}

When quantum Monte Carlo simulations are performed in frustrated magnetic systems, they generally suffer from the severe sign problem \cite{Sign_problem_QMC,Sign_problem_QMC2}. Well known representative of this class of systems are antiferromagnetic systems in a non-bipartite lattice, fermions in dimensions higher than one, anisotropic antiferromagnetic systems in the presence of an external magnetic field when simulated by the meron cluster algorithm (see discussion in section~2), among others. In all of these models, the Boltzmann factors associated with some possible configurations of the system in the path integral representation can be negative, and, therefore, cannot be interpreted as probabilities  \cite{Sign_Problem_Henelius}. This feature represents a difficult task to solve when defining the stochastic process. In general, the sign problem belongs to the non-polynomial class \cite{Sign_problem_NP} and should be solved case by case. Even if one could define positive transition probabilities and include the sign into the observable, so that accessibility and detailed balance are fulfilled, fluctuations of the sign in the sampled configurations would generate strong cancellations of the physical observables, which would make it impossible to measure them accurately using these numerical methods  \cite{Sign_Problem_Wiese}. In order to address this problem, the meron-cluster algorithm (MCA) has been originally proposed in \cite{Sign_Problem_Wiese,MCA}. This algorithm has an important feature of including clusters in the update procedure, which is essential to beat or at least to reduce critical slowing down \cite{Swendsen_Wang,Wolff}.

In this paper, we perform a numerical simulation of the quantum antiferromagnetic Heisenberg model (AFH) in both one and two dimensions in the presence of an external magnetic field by using the meron-cluster algorithm \cite{Sign_Problem_Wiese,MCA}. Extensive studies have been made in the spin 1/2 antiferromagnetic Heisenberg model, since this model can fairly well describe several physical systems, among which we mention undoped cuprates, which contains two dimensional {Cu}-{O} planes with antiferromagnetic interactions. Even more, novel quantum fluctuation in the {CuO}$_2$ planes, common in all these doped cuprates may be responsible for superconductivity at high temperature \cite{Anderson}, \cite{Berdnoz_Muller}. For the study of the low energy physics of these systems it is sufficient to consider the two dimensional AFH model, since there are many different quasi-two-dimensional materials with a square lattice structure, for example the cuprates {La}$_{2-x}${Sr}$_x${CuO}$_4$ and {YBa}$_2${Cu}$_3${O}$_{7-x}$. Although a rigorous study of doped materials is extremely hard to perform, the undoped precursors are antiferromagnetically ordered and can be modelled using the AFH model.

From the numerical point of view, the efficiency of the MCA as any cluster algorithm is based on the independency of the flips with probability 1/2 performed on the fly on each cluster built by the algorithm. This required feature would be destroyed for the present models if one would chose the quantization direction parallel to the direction of the external magnetic field. By choosing the quantization axis into a different direction, the {Z}(2) symmetry which allows all clusters to be independently flipped with probability 1/2, is restored, but a severe sign problem appears, which hampers the efficiency of the standard cluster algorithms that can be solved by the MCA.

It is worth mentioning that in \cite{Directed_loop} an alternative algorithm is proposed, which overcomes the above mention cluster update difficulty. In fact, in this reference, the concept of directed loop in stochastic series expansion and path-integral quantum Monte Carlo is introduced and successfully applied to the  $S = 1/2$ Heisenberg model in an external magnetic field by performing the quantization in the same direction of the magnetic field. The algorithm, which is technically demanding, works efficiently in all regions of the parameter space.
It represents a generalization of the loop-cluster algorithm and is based on a two-step method: first, a solution of the directed loop equations, which ensure a detailed balance, is constructed. It is further argued that the optimal solution should minimize the probability of backtracking in the loop construction and it, therefore, leads to a more efficient algorithm. In a second step, the concept of directed loops is applied to the path-integral Monte Carlo method formulated in imaginary time.  These ideas are corroborated by computing the integrated autocorrelation times as a function of the external magnetic field for a Heisenberg chain of 64 sites and inverse temperature  $\beta = 16$. The numerical results show that the algorithm works very efficiently, and should be more efficient than the worm algorithm \cite{WA_Prokofev,WA_Prokofev_2}. There is further raised a question whether the directed loop concept could be used to extend the meron notion to address the sign problem.

In the same spirit, the MCA \cite{Sign_Problem_Wiese} was proposed as a natural extension of the loop cluster algorithm (LCA) \cite{LCA_Wiese,LCA_Evertz}, applicable to systems with dynamical frustration, in which a severe sign problem could appear and spoil the efficiency of the LCA. Although the MCA was proposed over a decade ago and it represents a solution to the algorithm discontinuity in the sense discussed in \cite{Directed_loop}, its efficiency has so far not been measured properly. In order to address this task, we have implemented the meron-cluster algorithm \cite{Sign_Problem_Wiese,MCA} to the AFH model in the presence of an external magnetic field in the whole range of the magnetic field. We have combined the algorithm with the concept of improved estimators to measure the magnetization density, which corresponds to the order parameter in this model. The numerical results agree remarkably well with the corresponding values obtained in the article of Chandrasekharan et al. \cite{MCA} for quantum spin ladders, as well as with the ones obtained by using scaling behavior laws and extrapolation of them in the one- and two-dimensional models of reference \cite{AFH_Scaling_behavior}, as well as with the numerical results for the magnetization density in one and two dimensions, obtained by the alternative directed-loop quantum Monte Carlo method \cite{Directed_loop}.

In order to measure the statistical independence of the configurations generated by the MCA we made an effort to estimate the integrated autocorrelation time $\tau_\textrm{int}$ by using a binning method \cite{binning,binning_2}. A large statistic time series of the magnetization density, generated by the underlying stochastic dynamics defined by the meron-cluster algorithm, was analysed and an estimate for $\tau_\textrm{int}$ was obtained. This analysis gives a lower bound of Monte Carlo sweeps needed to accurately estimate expectation values as arithmetic means over the Markov chain.

There is another interesting physical feature involved in this kind of physical system, connected with the dynamical competition between the interactions which leads to a non-unique ground state at low temperature as in the quantum antiferromagnetic Heisenberg model in the presence of an external magnetic field. This property leads to a quantum phase transition (QPT) at zero temperature in such systems \cite{Sachdev}. In order to find and describe QPTs, different observables have been proposed, such as the entanglement of formation, the concurrence among others \cite{Wooters,Wooters_2}. We have studied the behaviour of the magnetization density and its first derivative for the AFH model and found numerical evidence of a second order QPT very close to the critical values $B_\textrm{c} = 2$ and $B_\textrm{c} = 4$, for the system in one and two spatial dimensions, which are consistent with the second order critical point conjectured in \cite{sl_qpt} and \cite{E. Bublitz-J. Ricardo de Sousa}, respectively.

\section{The Meron-cluster algorithm}

In this section we implement the MCA for the AFH model in the presence of an external magnetic field. This algorithm is an extension of the LCA, whose further technical details can be found in references \cite{Sign_Problem_Wiese,MCA,LCA_Wiese} and \cite{LCA_Evertz}. We consider a system of $1/2$ quantum spins defined on both a lattice of one dimension and on a square lattice, with site labels $x$, $y$ and equipped with periodic boundary conditions. The spin operators fulfil the SU(2) algebra of commutations:
\begin{equation}
\left[ \hat{S}_x^i, \hat{S}_y^j\right] = \ri \varepsilon_{ijk} \hat{S}_x^k \delta_{x,y}\, ,
\end{equation}
where $\hat{S}_x^i$ is the $i$ component of the spin operator defined on site with label $x$. The Hamilton operator
\begin{equation}
\hat{\mathcal{H}} = J\sum_{\langle x,y \rangle} \hat{\vec{S}}_x \cdot \hat{\vec{S}}_y - B \sum_x \hat{S}_x^1 = \sum_{\langle x,y \rangle} \hat{h}_{x,y} + \sum_x \hat{h}_{B_x}
\label{Hamiltonian}
\end{equation}
with:
\begin{equation}
\hat{h}_{x,y} = J \hat{\vec{S}}_x \cdot \hat{\vec{S}}_y \,, \qquad
\hat{h}_{B_x} = -B \hat{S}_x^1
\end{equation}
describes the interaction between nearest neighbours on the lattice, where $J > 0$ is the coupling constant of the antiferromagnetic interaction and $B$ represents an external constant magnetic field.

We will compute the partition function $Z$ of the system written as a path integral. The path integral representation is obtained by decomposing the Hamiltonian into $(2d+1)$ $\hat{H}_i$ pieces, where $d = 1,2$ is the system dimension and by using the Trotter-Suzuki formula \cite{Trotter,Suzuki}. For $d = 1$ we choose:
\begin{equation}
\hat{H}_1 = \sum_{x  \text{ even}}
 \hat{h}_{x,x+\hat{\textbf{1}}}\,, \qquad \hat{H}_2 = \sum_{x \ \text{odd}} \hat{h}_{x,x+\hat{\textbf{1}}}\,, \qquad
\hat{H}_B = \sum_x \hat{h}_{B_x}
\label{MCA_Hamiltonian_decomp_1d}
\end{equation}
and for $d = 2$:
\begin{align}
\hat{H}_1 &= \sum_{\substack{x=(m,n) \\ m+n \ \text{even}}}
 \hat{h}_{x,x+\hat{\textbf{2}}}\,, \qquad \hat{H}_2 = \sum_{\substack{x=(m,n) \\ m+n \ \text{even}}} \hat{h}_{x,x+\hat{\textbf{1}}}\,, \qquad
\hat{H}_3 = \sum_{\substack{x=(m,n) \\ m+n \ \text{odd}}} \hat{h}_{x,x+\hat{\textbf{2}}}\,, \qquad \nonumber \\
\hat{H}_4 &= \sum_{\substack{x=(m,n) \\ m+n \ \text{odd}}} \hat{h}_{x,x+\hat{\textbf{1}}} \,, \qquad \hat{H}_B = \sum_x \hat{h}_{B_x}\,.
\label{MCA_Hamiltonian_decomp_2d}
\end{align}

Then, the partition function can be written as
\begin{equation}
Z = \lim_{N \rightarrow \infty} \textrm{Tr} \left\{ \left[ \prod_{i = 1}^{2d+1}\exp(-\epsilon \hat{H}_i) \right]^N \right\}.
\end{equation}

 The important point to note here is that all the individual terms present in $\hat{H}_i$ commute with each other, but two different $\hat{H}_i$ and $\hat{H}_j$ do not commute. This allows us to introduce $(2d+1)N$ identities between each exponential factor in order to derive the path integral. Here, $\epsilon= \beta / N$ represents the elemental lattice spacing in the extra dimension of length $(2d+1)N$.

Once these formal manipulations have been performed, one obtains a $(d+1)$-dimensional lattice of classical Ising-type of spins. Using the eigenstates $\left| \uparrow \right \rangle$ and $\left| \downarrow \right>$ of the $\hat{\sigma}^3$ operator, the transfer matrices turn out to be:

\begin{equation}
\hat{\mathcal{T}}_1 \equiv \re^{-\epsilon \hat{h}_{x,y}}=\begin{bmatrix} \cal{A} & 0 & 0 &0 \\0 & \cal{B} & \cal{C} & 0 \\0 & \cal{C} & \cal{B} & 0 \\0 & 0 &0 &\cal{A} \end{bmatrix}\quad \begin{matrix} \left|\uparrow \hspace{0.1 cm} \uparrow \right>\\\left|\uparrow \hspace{0.1 cm} \downarrow \right> \\ \left|\downarrow \hspace{0.1 cm} \uparrow \right> \\\left|\downarrow \hspace{0.1 cm} \downarrow \right> \end{matrix}\,,
\end{equation}
\begin{equation}
\hat{\mathcal{T}}_B \equiv \re^{-\epsilon \hat{h}_{B_x}}=\begin{bmatrix} \cal{E} & \cal{F}\\ \cal{F} & \cal{E} \end{bmatrix}\quad \begin{matrix} \left|\uparrow \right>\\\left|\downarrow \right> \end{matrix}\,.
\end{equation}

The non-zero matrix elements of $\hat{\mathcal{T}}_1$ (up to an irrelevant overall pre-factor $\re^{\epsilon J/4}$) are:
\begin{align}
\label{Transfer_matrix1}
\mathcal{A} &= \left< \uparrow \hspace{0.1 cm}\uparrow \right|\exp{(-\epsilon \hat{h}_{x,y})}\left| \uparrow \hspace{0.1 cm} \uparrow \right> =  \left< \downarrow \hspace{0.1 cm}\downarrow \right|\exp{(-\epsilon \hat{h}_{x,y})}\left| \downarrow \hspace{0.1 cm} \downarrow \right> = \exp(-\epsilon J/2), \nonumber \\
\mathcal{B} &= \left< \uparrow \hspace{0.1 cm}\downarrow \right|\exp{(-\epsilon \hat{h}_{x,y})}\left| \uparrow \hspace{0.1 cm} \downarrow \right> = \left< \downarrow \hspace{0.1 cm}\uparrow \right|\exp{(-\epsilon \hat{h}_{x,y})}\left| \downarrow \hspace{0.1 cm} \uparrow \right> = \cosh{(\epsilon J/2)}, \\
\mathcal{C} &= \left< \uparrow \hspace{0.1 cm}\downarrow \right|\exp{(-\epsilon \hat{h}_{x,y})}\left| \downarrow \hspace{0.1 cm} \uparrow \right> = \left< \downarrow \hspace{0.1 cm}\uparrow \right|\exp{(-\epsilon \hat{h}_{x,y})}\left| \uparrow \hspace{0.1 cm} \downarrow \right> = - \sinh{(\epsilon J/2)}, \nonumber
\end{align}
while the corresponding non-trivial matrix elements of $\hat{\mathcal{T}}_B$ are:
\begin{align}
\label{Transfer_matrix2}
\mathcal{E} &= \left< \uparrow \right| \exp{(- \epsilon \hat{h}_{B_x})} \left| \uparrow \right> = \left< \downarrow \right| \exp{(- \epsilon \hat{h}_{B_x})} \left| \downarrow \right> = \cosh{(\epsilon B/2)}, \\
\mathcal{F} &= \left< \uparrow \right| \exp{(- \epsilon \hat{h}_{B_x})} \left| \downarrow \right> = \left< \downarrow \right| \exp{(- \epsilon \hat{h}_{B_x})} \left| \uparrow \right> = \sinh{(\epsilon B/2)}, \nonumber
\end{align}
where in these equations, the bra vectors $ \left< \uparrow \right|$ and $ \left< \downarrow \right|$ are the dual eigenstates of $\hat{\sigma}^3$ defined on the time slice $t$, and the ket vectors $\left| \uparrow \right>$ and $\left| \downarrow \right>$
 are the $\hat{\sigma}^3$ eigenstates defined on the time slice $t + 1$ of the $d + 1$ dimensional lattice.

At this point it is worth mentioning that we have chosen the quantization direction perpendicular to the one of the magnetic field, so that one can still update whole spin clusters with probability 1/2. This choice is essential because a flip probability depending on the value of the magnetic field would rarely lead to possible flips of magnetized clusters for strong or weak fields, which would destroy the efficiency of most of cluster algorithms.

In the path integral representation of the partition function, the non-zero elements of the transfer matrices which are $\mathcal{A}$, $\mathcal{B}$, $\mathcal{C}$, $\mathcal{E}$ and $\mathcal{F}$ (known as plaquettes) define the allowed spin-configurations of the auxiliary lattice of $d+1$ dimensions. From equation (\ref{Transfer_matrix1}) and (\ref{Transfer_matrix2}) it follows that a sign problem arises for $J > 0$ and $ B < 0$, respectively. The minus sign appearing in $\mathcal{F}$ can be easily avoided by making the choice $B > 0$ which does not affect the physical contents described by the Hamiltonian. Nevertheless, for an anti-ferromagnetic interaction, $J$ is necessarily positive, which leads to a minus sign in the matrix element $\mathcal{C}$ and prevents interpreting this term as a Boltzmann factor (or probability) in the simulation process.

In order to get rid of the sign problem, it is useful first to rewrite the matrix elements as a product of the sign of the plaquettes, times the Boltzmann's factor associated:
\begin{align}
\mathcal{A} &= \left< \uparrow \hspace{0.1 cm}\uparrow \right|\exp{({-\epsilon \hat{h}_{x,y}})}\left| \uparrow \hspace{0.1 cm} \uparrow \right> = \text{Sign}\left[ \left< \uparrow \hspace{0.1 cm}\uparrow \right|\exp({-\epsilon \hat{h}_{x,y}})\left| \uparrow \hspace{0.1 cm} \uparrow \right> \right] \exp\left\{- \mathcal{S} \left[ \left< \uparrow \hspace{0.1 cm}\uparrow \right|\exp{(-\epsilon \hat{h}_{x,y})}\left| \uparrow \hspace{0.1 cm} \uparrow \right> \right] \right\}, \nonumber \\
\mathcal{B} &= \left< \uparrow \hspace{0.1 cm}\downarrow \right|\exp({-\epsilon \hat{h}_{x,y}})\left| \uparrow \hspace{0.1 cm} \downarrow \right> = \text{Sign}\left[ \left< \uparrow \hspace{0.1 cm}\downarrow \right|\exp({-\epsilon \hat{h}_{x,y}})\left| \uparrow \hspace{0.1 cm} \downarrow \right> \right] \exp\left\{- \mathcal{S} \left[ \left< \uparrow \hspace{0.1 cm}\downarrow \right|\exp{(-\epsilon \hat{h}_{x,y})}\left| \uparrow \hspace{0.1 cm} \downarrow \right> \right] \right\},  \\
\mathcal{C} &= \left< \uparrow \hspace{0.1 cm}\downarrow \right|\exp({-\epsilon \hat{h}_{x,y}})\left| \downarrow \hspace{0.1 cm} \uparrow \right> = \text{Sign}\left[ \left< \uparrow \hspace{0.1 cm}\downarrow \right|\exp({-\epsilon \hat{h}_{x,y}})\left| \downarrow \hspace{0.1 cm} \uparrow \right> \right] \exp\left\{- \mathcal{S} \left[ \left< \uparrow \hspace{0.1 cm}\downarrow \right|\exp{(-\epsilon \hat{h}_{x,y})}\left| \downarrow \hspace{0.1 cm} \uparrow \right> \right] \right\} \nonumber
\label{aver}
\end{align}
and
\begin{align}
\mathcal{E} &= \left< \uparrow \right| \exp{(- \epsilon \hat{h}_{B_x})} \left| \uparrow \right> = \text{Sign}\left[ \left< \uparrow \right| \exp{(- \epsilon \hat{h}_{B_x})} \left| \uparrow \right> \right] \exp\left\{- \mathcal{S} \left[ \left< \uparrow \right| \exp{(- \epsilon \hat{h}_{B_x})} \left| \uparrow \right> \right] \right\},
 \\
\mathcal{F} &= \left< \uparrow \right| \exp{(- \epsilon \hat{h}_{B_x})} \left| \downarrow \right> = \text{Sign}\left[ \left< \uparrow \right| \exp{(- \epsilon \hat{h}_{B_x})} \left| \downarrow \right> \right] \exp\left\{- \mathcal{S} \left[ \left< \uparrow \right| \exp{(- \epsilon \hat{h}_{B_x})} \left| \downarrow \right> \right] \right\}, \nonumber
\end{align}
where $\textrm{Sign}[x]$ is the usual sign function. Thus, the signs of each plaquette are $\text{Sign}[\mathcal{A}]=\text{Sign}[\mathcal{B}]=\text{Sign}[\mathcal{E}] = 1$, $\text{Sign}[\mathcal{C}] = - \text{Sign}[J] $ and $\text{Sign}[\mathcal{F}] = \text{Sign}[B] $, and the corresponding Boltzmann's factors turn to be:
\begin{align}
\re^{-\mathcal{S}[\mathcal{A}]} &= \exp(-\epsilon J/2), \nonumber\\
\re^{-\mathcal{S}[\mathcal{B}]} &= \cosh(\epsilon J/2), \nonumber\\
\re^{-\mathcal{S}[\mathcal{C}]} &= \sinh(\epsilon |J|/2),\\
\re^{-\mathcal{S}[\mathcal{E}]} &= \cosh(\epsilon B/2), \nonumber\\
\re^{-\mathcal{S}[\mathcal{F}]} &= \sinh(\epsilon |B|/2).\nonumber
\label{Boltzmann_factor}
\end{align}

Using the sign values of these plaquettes we can define the global sign of a lattice configuration as the product of the signs of the plaquettes which the configuration is made of. It follows that the possible values of the sign of a configuration $[s]$ in the auxiliary lattice can be Sign$[s] = 1$ or Sign$[s] = -1$. The partition function can also be written as:
\begin{equation}
Z= \sum_{[s]} \text{Sign}[s] \exp{(-\mathcal{S}[s] )}.
\end{equation}
This equation represents the partition function of the original quantum spin system defined on a $d$-dimensional lattice written as a path integral over configurations $[s]$ of the auxiliary $d+1$ dimensional lattice with an associated Boltzmann weight $\exp{(-\mathcal{S}[s])}$ and $\text{Sign}[s] = \pm 1$.

We now briefly explain how the meron-cluster algorithm gets rid of the sign problem. The main idea is to compute the expectation value of an observable $\langle \hat{O} \rangle$ in the original system by using the modified system with action $\mathcal{S}[s]$, which does not suffer from the sign problem:
\begin{equation}
\langle \hat{O} \rangle = \frac{1}{Z}\sum_{[s]} O[s] \text{Sign}[s]\exp{(-\mathcal{S}[s])} = \frac{\langle \hat{O} \text{Sign} \rangle }{\left< \text{Sign} \right>}\,.
\end{equation}
It follows that the original expectation value $\langle \hat{O} \rangle$ is calculated as the ratio of two modified expectation values $\langle \hat{O} \text{Sign} \rangle$ and $\left< \text{Sign} \right>$ which are exponentially small in both the inverse temperature $\beta$ and the volume system as it was shown in \cite{Sign_Problem_Wiese}. Due to the strong cancellations coming from the sign fluctuations, both terms appearing on this ratio are very small compared to their statistical errors and, therefore, it is in practice impossible to measure them accurately. The solution to this apparent problem involves a two-step procedure: in a first step, the algorithm matches any contribution with one sign with another paired contribution with the opposite sign, which will not give a net contribution to the observable. Only a few unmatched relevant contributions remain. In a second step, the algorithm discard these paired contributions by using a Metropolis decision, and includes only those contributions to the observables that are relevant (see discussion of the improved estimators in the next section). This step is called reweighting, and aims at suppressing multi-meron configurations, which should not be sampled in the stochastic process because they do not contribute to the physical observables. See in the next section a discussion on the multi-meron concept.

\begin{figure}[!t]
\begin{center}
	\begin{tikzpicture}[scale = .8]
	\draw [gray, fill=lightgray] (0-.5,0) -- (1-.5,0) -- (1-.5,1) -- (0-.5,1)--(0-.5,0);\draw [gray, fill=lightgray] (0+1.5,0) -- (1+1.5,0) -- (1+1.5,1) -- (0+1.5,1)--(0+1.5,0);
	\draw[black, line width=3.5pt] (0+1.5,0)--(0+1.5,1); \draw[black, line width=3.5pt] (1+1.5,0)--(1+1.5,1);
	\draw[black, fill=white] (0-.5,0) circle (1mm);\draw[black, fill=white] (1-.5,0) circle (1mm);\draw[black, fill=white] (1-.5,1) circle (1mm);\draw[black, fill=white] (0-.5,1) circle (1mm);
	\draw[black, fill=white] (0+1.5,0) circle (1mm);\draw[black, fill=white] (1+1.5,0) circle (1mm);\draw[black, fill=white] (1+1.5,1) circle (1mm);\draw[black, fill=white] (0+1.5,1) circle (1mm);
\draw(-1,0.3)node{$\mathcal{A}$};
\draw(2.0,1.3)node{$P_{||}^\mathcal{A}$};
	\draw [gray, fill=lightgray] (0-.5,0-2) -- (1-.5,0-2) -- (1-.5,1-2) -- (0-.5,1-2)--(0-.5,0-2);\draw [gray, fill=lightgray] (0+1.5,0-2) -- (1+1.5,0-2) -- (1+1.5,1-2) -- (0+1.5,1-2)--(0+1.5,0-2);\draw [gray, fill=lightgray] (0+3,0-2) -- (1+3,0-2) -- (1+3,1-2) -- (0+3,1-2);\draw[black, line width=3.5pt] (0+1.5,0-2)--(0+1.5,1-2); \draw[black, line width=3.5pt] (1+1.5,0-2)--(1+1.5,1-2);
	\draw[black, line width=3.5pt] (0+3,0-2)--(1+3,0-2);\draw[black, line width=3.5pt] (0+3,1-2)--(1+3,1-2);
	\draw[black, fill=white] (0-.5,0-2) circle (1mm);\draw[black, fill=black] (1-.5,0-2) circle (1mm);\draw[black, fill=black] (1-.5,1-2) circle (1mm);\draw[black, fill=white] (0-.5,1-2) circle (1mm);
	\draw[black, fill=white] (0+1.5,0-2) circle (1mm);\draw[black, fill=black] (1+1.5,0-2) circle (1mm);\draw[black, fill=black] (1+1.5,1-2) circle (1mm);\draw[black, fill=white] (0+1.5,1-2) circle (1mm);
		\draw[black, fill=white] (0+3,0-2) circle (1mm);\draw[black, fill=black] (1+3,0-2) circle (1mm);\draw[black, fill=black] (1+3,1-2) circle (1mm);\draw[black, fill=white] (0+3,1-2) circle (1mm);
\draw(-1,0.3-2)node{$\mathcal{B}$};
\draw(2.0,1.3-2)node{$P_{||}^\mathcal{B}$};\draw(2.0+1.5,1.3-2)node{$P_{=}^\mathcal{B}$};
	\draw [gray, fill=lightgray] (0-.5,0-4) -- (1-.5,0-4) -- (1-.5,1-4) -- (0-.5,1-4)--(0-.5,0-4);\draw [gray, fill=lightgray] (0+1.5,0-4) -- (1+1.5,0-4) -- (1+1.5,1-4) -- (0+1.5,1-4)--(0+1.5,0-4);
	\draw[black, line width=3.5pt] (0+1.5,0-4)--(1+1.5,0-4);\draw[black, line width=3.5pt] (1+1.5,1-4)--(0+1.5,1-4);
	\draw[black, fill=white] (0-.5,0-4) circle (1mm);\draw[black, fill=black] (1-.5,0-4) circle (1mm);\draw[black, fill=white] (1-.5,1-4) circle (1mm);\draw[black, fill=black] (0-.5,1-4) circle (1mm);
	\draw[black, fill=white] (0+1.5,0-4) circle (1mm);\draw[black, fill=black] (1+1.5,0-4) circle (1mm);\draw[black, fill=white] (1+1.5,1-4) circle (1mm);\draw[black, fill=black] (0+1.5,1-4) circle (1mm);
\draw(-1,0.3-4)node{$\mathcal{C}$};
\draw(2.0,1.3-4)node{$P_{=}^\mathcal{C}$};
\draw[dashed,line width=0.5pt] (5-0.2,1.5)--(5-0.2,-4.2);
\draw[lightgray, line width=4.0pt] (6,0)--(6,1); \draw[lightgray, line width=4.0pt] (6+1.6,0)--(6+1.6,1);\draw[black, line width=3.5pt] (6+1,0)--(6+1,1);
\draw[black, fill=white] (6,0) circle (1mm);\draw[black, fill=white] (6,1) circle (1mm);
\draw[black, fill=white] (6+1,0) circle (1mm);\draw[black, fill=white] (6+1,1) circle (1mm);
\draw[black, fill=white] (6+1.6,0) circle (1mm);\draw[black, fill=white] (6+1.6,1) circle (1mm);
\draw(-1+.5+6,0.3)node{$\mathcal{E}$};
\draw(1.0+6,1.4)node{$P_{|}^\mathcal{E}$};\draw(1.7+6,1.4)node{$P_{:}^\mathcal{E}$};
\draw[lightgray, line width=4.0pt] (6,0-8+6)--(6,1-8+6); \draw[lightgray, line width=4.0pt] (6+1,0-8+6)--(6+1,1-8+6);
\draw[black, fill=white] (6,0-8+6) circle (1mm);\draw[black, fill=black] (6,1-8+6) circle (1mm);
\draw[black, fill=white] (6+1,0-8+6) circle (1mm);\draw[black, fill=black] (6+1,1-8+6) circle (1mm);
\draw(6+-1+.5,0.3-8+6)node{$\mathcal{F}$};
\draw(6+1.0,1.4-8+6)node{$P_{:}^\mathcal{F}$};
  \end{tikzpicture}
\end{center}
\vspace{-5mm}
\caption{Suitable break-ups associated to each plaquette that appear in the left of each row and correspond to the non-trivial elements of the transfer matrices $\hat{\mathcal{T}}_1$  and $\hat{\mathcal{T}}_B$, respectively. The filled circles represent spins 1 and empty circles represent spins $-1$. The break-ups represented by bold lines are displayed with their corresponding probabilities.}
\label{MCA_Breakups}
\end{figure}
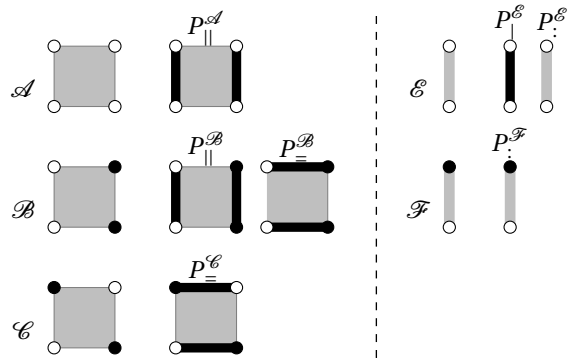

In the modified system without the minus signs it is possible to define a cluster algorithm by introducing break-ups that correspond to a suitable type of bonds between the spins within the plaquette, which is currently considered in the dynamical update procedure defined by the algorithm. In this way, a Markov chain is built which fulfils a detailed balance and the micro-causality condition in a similar way as it was performed with the loop cluster algorithm \cite{LCA_Wiese,LCA_Evertz}. In fact, in the modified system, the weight is always positive and, therefore, can be interpreted as Boltzmann's factors, which leads to the break-ups displayed in figure~\ref{MCA_Breakups}.

The probabilities are defined by: $P_{||}^\mathcal{A} = P_{=}^\mathcal{C} = P_{:}^\mathcal{F} = 1$, $P_{||}^\mathcal{B} =  \exp{(- \epsilon J/2)}/\cosh{(\epsilon J/2)}$ $= 1 - P_{=}^\mathcal{B}$ and $ P_{:}^\mathcal{E} = 1 - P_{|}^\mathcal{E} = \tanh{(\epsilon |B|/2)} $. By joining together the resulting break-ups on each plaquette, different strings are produced, which we will call clusters. Due to the ``$:$'' break-up defined for the $\mathcal{E}$ and $\mathcal{F}$ plaquettes, the clusters will not necessarily correspond to closed loops. Nevertheless, each spin $s$ of the $d+1$ dimensional lattice will belong to only one cluster and each cluster will be updated independently with probability $1/2$.

\section{Improved estimators}

The sign of a configuration $\text{Sign}[s]$ was defined as the product of the signs of the plaquettes which it is made of. This distinction can be translated to the clusters themselves by defining a meron cluster as a cluster whose flip would change $\text{Sign}[s]$. This definition leads to the distinction of the configuration space between the zero-meron sector, which consists of only non-meron clusters, and the meron sector, which consists of at least one meron cluster. Multi-meron configurations are configurations with more than one meron cluster. Using the meron algorithm allows us to gain an exponential factor in statistic, and moreover, if it is combined with the improved estimators techniques, it becomes a powerful tool to accurately compute the observables in reachable computational times even for relative large lattices. Improved estimators, as their name points out, are techniques to compute the estimators for the observables in an improved manner in such a way to speed up the whole numerical simulation, which can be achieved by measuring global properties of the clusters defined in each configuration of the system generated in the Markov chain (which can be done on the fly, when clusters are built); instead of measuring local parameters or the properties that are more costly in computational time.

In particular, magnetization is an interesting physical observable. Due to symmetry properties of the physical system described by the Hamiltonian of equation (\ref{Hamiltonian}), the only non-trivial component of the magnetization is the one parallel to the external magnetic field. The average of this component can be very efficiently measured by computing the temporal winding number of the closed-loops which result from joining the open string clusters \cite{MCA}:
\begin{equation}
\langle \hat{M}^1 \rangle = \frac{1}{2 \langle \delta_{\mathcal{N},0} \rangle} \langle \delta_{\mathcal{N},0} \sum_j W_j \rangle,
\label{M_winding}
\end{equation}
where $W_j$ is the temporal winding number associated to the closed loop $j$. If the loop is not made by open string clusters, then it does not contribute to the observable  $W = 0. \hspace{0.2 cm} \mathcal{N}$ corresponds to the number of meron strings appearing in the considered configuration $[s]$. The kronecker's delta constraints the configuration space to only the zero-meron sector, from which the magnetization gets non-vanishing contributions.

Since the operators $\hat{S}_x^1 $ are not diagonal in the base used for quantization, it is not straightforward to understand how can  $\langle \hat{M}^1 \rangle$  be measured by using the temporal winding number of the joined open string clusters generated through the numerical simulation. In references \cite{MCA,Brower_Wiese,Sandvik_comp_Quant}, the authors show how non-diagonal operators can be measured  through the Markov chain, the reader can see these references for further details. Nevertheless, it is straightforward to understand how the winding number can be used to compute this magnetization if the quantization was made in the direction of the external magnetic field. Although, as it was explained above, the freedom of independently flipping each cluster by a coin decision is lost in this case. With the purpose of helping the reader to understand how topologycal properties of clusters, as the winding number, can be used to compute the magnetization, we include herein below a brief explanation for a diagonal operator case.

By putting the direction of the magnetic field in the third direction (the same of the quantization axis), the sign problem disappears and the non-zero contribution of the magnetization is given by the average of $\hat{M}^3$. The matrix representation of $\hat{S}_x^3$ is thus diagonal, and it holds:
\begin{equation}
\langle \hat{M}^3 \rangle = \lim_{N \rightarrow \infty} \frac{1}{Z} \textrm{Tr} \left\{ \hat{M}^3 \left[ \prod_{i = 1}^{2d+1}\exp(-\epsilon \hat{H}_i) \right]^N \right\}  =\frac{1}{Z} \sum_{[s]} M[s] \exp{(-\mathcal{S}[s])}.
\end{equation}
Since the cluster algorithm produces the configuration $[s]$ with statistical weight:
\begin{equation}
\mathcal{P}[s] = \frac{1}{Z}\exp{(-\mathcal{S}[s])}
\end{equation}
it follows:
\begin{equation}
\langle \hat{M}^3 \rangle = \langle M[s] \rangle.
\end{equation}
Due the trace invariance, for a given configuration $[s]$, we can choose $M[s]$ as the sum of the spin values at any time slice $t$ in the auxiliary lattice of $d+1$ dimensions. Or even better, we may compute it as the sum of all spin values of the $[s]$ configuration, with a corresponding normalization factor in order to avoid an overcount. By this last choice, the sum can be  very efficiently realized through clusters, since each spin will belong to only one cluster:
\begin{equation}
M[s] = \frac{1}{2 L_t}\sum_i m_{C_i}\,,
\end{equation}
where $m_{C_i}$ corresponds to the sum of the spin values of the i-th cluster, and $L_t = (2d+1)N$ represents the length of the lattice in the temporal direction. As shown in figure~\ref{MCA_Breakups}, only equal spin values can be linked by bonds in the temporal direction and only different spin values can be linked by bonds in spatial direction. Moreover, it is worthwhile  mentioning as $\mathcal{F} = 0$ in this case, that all clusters correspond to closed loops. If we assume that a cluster grows in the positive temporal direction only if it connects positive spin values $s = 1$ and grows in the negative temporal direction if it connects negative spins values $s = -1$, it follows $m_{C_i} = L_t W_i$, where $W_i$ corresponds to the temporal winding number of the cluster $C_i$, and one obtains finally:
\begin{equation}
\langle \hat{M}^3 \rangle = \frac{1}{2}\langle \sum_j W_j \rangle
\end{equation}
which is the analog result of equation (\ref{M_winding}), but for the quantization axis chosen parallel to the direction of the external magnetic field. In the next section we present numerical results for the magnetization of the considered models using equation (\ref{M_winding}), obtained by the average performed through the sampled configurations generated by the MCA.

Since we are interested in the magnetization, it is in principle unnecessary to generate configurations that contain meron-clusters. This is a crucial point of the meron-cluster algorithm as it allows us to gain an exponential factor in statistics by restricting the simulation to only the zero-meron sector, which is exponentially small compared to the whole configuration space. This could be done by associating a zero Boltzmann weight to any configuration with at least one meron-cluster. In spite of this enormous statistical advantage, complete avoidance of this meron-sector can lead to a slow-down of the algorithm. For that reason, it turned out advantageous to allow some meron configurations which do not contribute to the magnetization but speed up the stochastic dynamics by reducing the autocorrelation time. This step was performed by introducing an additional Boltzmann-type factor $q<1$ for each meron, so that if a new configuration with $\mathcal{N}'$ merons is proposed starting from a configuration containing $\mathcal{N}$ merons, the new configuration is accepted with a conditional probability $p=\textrm{min}[1,q^{(\mathcal{N}'-\mathcal{N})}]$. As it was shown in \cite{MCA}, adding this extra Boltzmann factor, in order to suppress multi-meron clusters for the re-weighting, only modifies the algorithm but the physics remains the same.

In order to ensure that the Markov chain generated by the algorithm should be ergodic, one should first show that any allowed configuration can be reached starting from a reference configuration. Moreover, this process should be reversible. It turns out that any allowed configuration for the auxiliary lattice in $d+1$ dimensions is formed by a combination of the allowed plaquettes. From this point of view, starting from the reference configuration, any configuration can be reached by sequentially changing a set of plaquettes using the breakups defined in figure~\ref{MCA_Breakups}, with associated probabilities defined at the end of section~2. As this process is microscopically reversible (the probabilities are non-zero in both directions), one can reach the reference configuration starting from an arbitrary configuration. This argument suggests that ergodicity is fulfilled.


\section{Numerical results}

\begin{figure}[!b]
\centering
\includegraphics[width=0.55\linewidth]{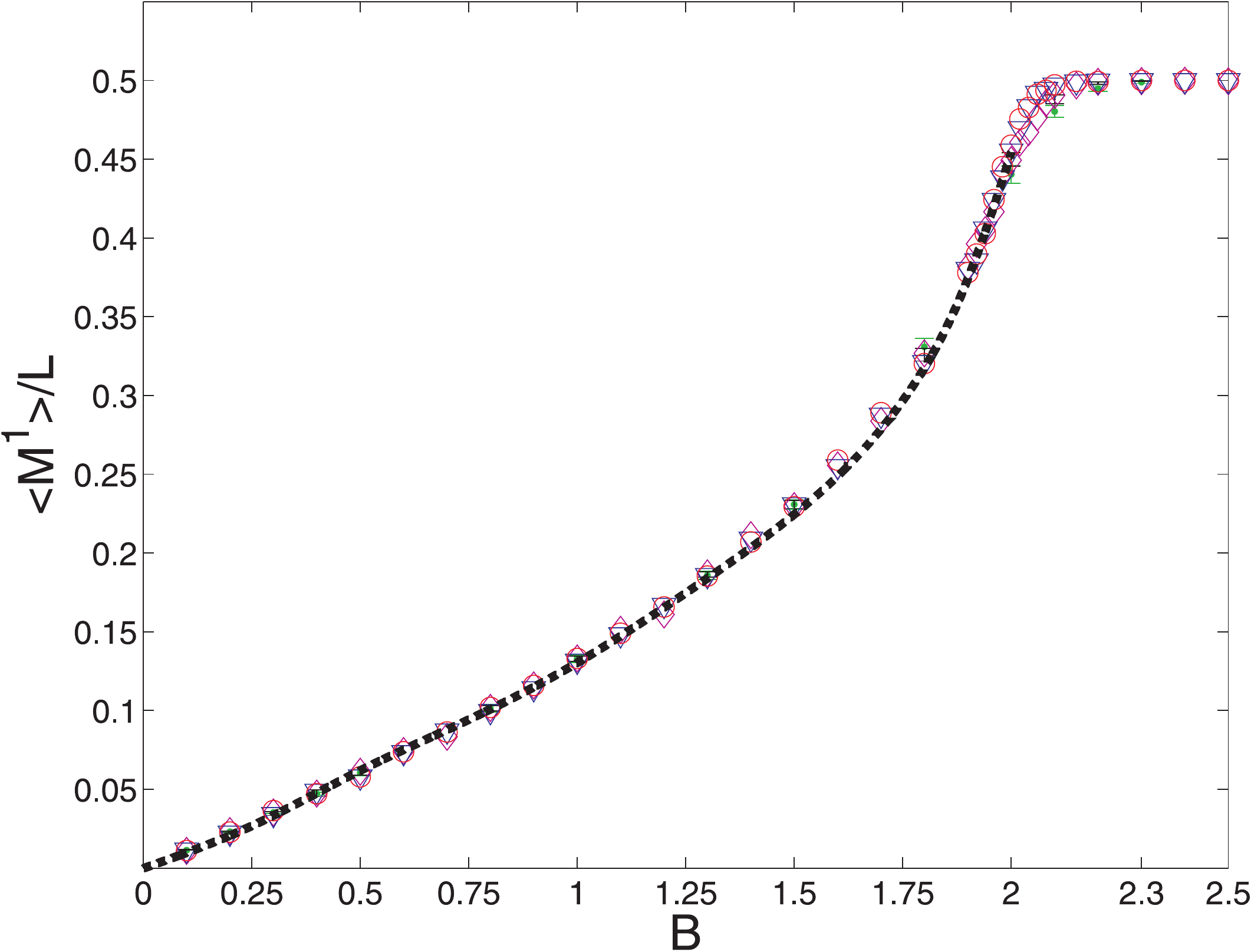}
\caption{(Color online) The magnetization density is plotted as a function of the external magnetic field $B$ for the AFH model in $d=1$ dimension. The dot and cross data correspond to a ring of length $L = 60$  for inverse temperature values $\beta = 15$ and $\beta = 20$, respectively; the diamond data correspond to $L = 80$ for inverse temperature $\beta = 20$, the inverted triangle data correspond to $L = 160$ for inverse temperature $\beta = 30$. and the empty circle data correspond to $L = 240$ at inverse temperature $\beta = 30$. All these simulations were performed using $\epsilon = 0.05$. The dot line corresponds to Fabricius et al. data reported in \cite{AFH_Scaling_behavior}, which were obtained based on the scaling behaviour and on the Bethe Ansatz for the magnetization in the thermodynamic limit $L \rightarrow \infty$ and $T = 0$.}
\label{mag1dAFH}
\end{figure}

Now we present numerical results obtained by using the meron-cluster algorithm briefly explained in the last two sections. We studied the anti-ferromagnetic Heisenberg model defined both in one- and in two-spatial dimensions. The spin-spin interaction constant was set equal to $J=1$, while the coupling $B$ for the external magnetic field was varied in the physical region $0<B<6$ of interest.
As the first application, we studied the one-dimensional Heisenberg model in the low-temperature region. In figure~\ref{mag1dAFH} the magnetization density is shown as a function of the external magnetic field for four different lattice sizes $L = 60, 80, 160$ and $240$ and for three values of the inverse of the temperature $\beta$, 15, 20 and 30. As a comparison, we also include the results obtained in \cite{AFH_Scaling_behavior}, which are based on exact diagonalization on small lattice sizes, guessed scaling relations and on the Bethe ansatz to extract the thermodynamic limit of the magnetization at zero temperature. Both results very accurately agree in the whole range of the external magnetic field $0<B<2$ which shows that indeed the finite size effects are not really important in one dimension. In order to estimate the errors, we have computed the statistical errors for a small lattice size at $\beta = 15$ and 20, and for different magnetic field values. It turns out that they are rather small, typically of a few percent up to 3 percent, and cannot be distinguish in figure~\ref{mag1dAFH}. As the external magnetic field goes to zero, the magnetization density goes also to zero, in agreement with the Mermin-Wagner theorem. Close to the threshold value $B=2$, the magnetization density saturates to its maximum value $1/2$ which corresponds to a perfect alignment of the spins. These results show that the meron-cluster algorithm represents a powerful tool to simulate frustrated magnetic systems close to zero temperature and encourages one to use it to study the structure of quantum phase transitions.

As it is well known, the antiferromagnetic Heisenberg (AFH) spin chain at zero temperature in the presence of an external magnetic field has a quantum phase transition (QPT) at the critical value $B_\textrm{c} = 2 $. At this critical value, the correlation length diverges and some physical quantities like the first derivative of the entanglement entropy and the susceptibility are singular which are general features of QPT \cite{Sachdev}. Above this critical value, the ground state becomes ferromagnetic in the thermodynamic limit, while below it, the physics is governed by magnons, and the antiferromagnetic order is preferred by the system.

In \cite{AFH_Scaling_behavior} it was shown that the magnetization curve develops a square root singularity as the magnetic field approaches its critical value $B_\textrm{c}= 2$.
Quantum entanglement as well as the concurrence have been widely used to find QPT in quantum systems \cite{Wooters,Wooters_2}. In \cite{sl_qpt}, the same physical system is studied by using exact diagonalization combined with the infinite time-evolving block decimation technique to compute the entanglement, the ground state energy and the magnetization as a function of the external magnetic field. It is shown that the quantum entanglement is sensitive to the subtle changing ground state and that it can be used to describe the magnetization and the QPT. In particular, they show a numerical evidence supporting that the magnetization displays a square root singularity as $B \rightarrow 2$, and its derivative becomes discontinuous at $B_\textrm{c} = 2$, beyond it the system becomes ferromagnetic. It is further conjectured that this point corresponds to the second order QPT.

\begin{figure}[!t]
\centering
\includegraphics[width=0.98\linewidth]{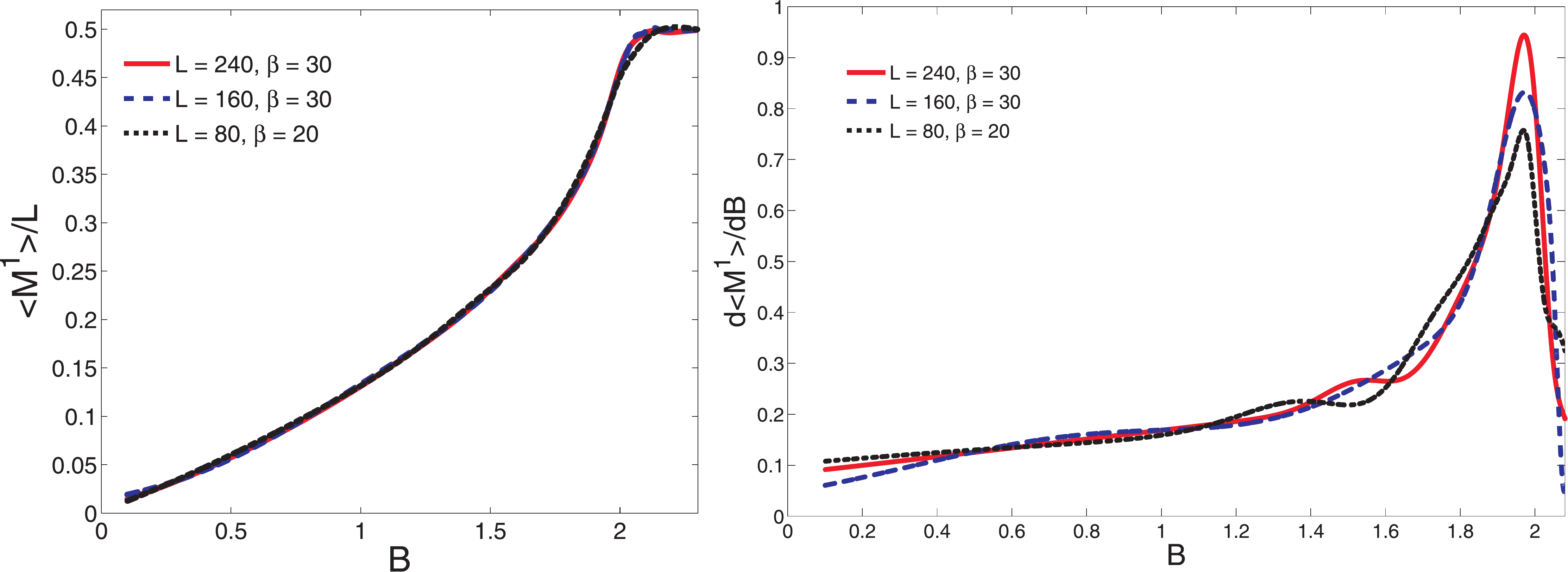}
\hspace{4cm}(a) \hspace{7cm} (b)
\caption{(Color online) (a) Fit of the numerical data plotted in figure~\ref{mag1dAFH}. (b) First-order derivative of the magnetization density as a function of the external magnetic field $B$ for the AFH spin chain. It shows the evidence for discontinuity close the critical value $B_\textrm{c} = 1.97$, for a large system size L and in the low temperature region.}
\label{mag_derivativeAFH1D}
\end{figure}

In figure~\ref{mag_derivativeAFH1D}~(b), the first derivative of the magnetization density is shown for the AFH spin chain for $L=80, 160$ and $240$ and inverse temperature $\beta=20$ and $\beta=30$, which was obtained by performing a high precision ($R^2 = 0.9999$) fit of the magnetization density with a suitable function, and computing its analytical first derivative [this fit is explicitely shown in figure~\ref{mag_derivativeAFH1D}~(a)]. In the very low temperature region, as the system size increases, the shape of the magnetization density collapses into a single curve displayed in figure~\ref{mag1dAFH}. Therefore, the behaviour of its first derivative, allows us to support the existence of a discontinuity at about $B_\textrm{c} = 1.97$ as $L$ goes to infinity and at zero temperature, which is very close to the critical value $B_\textrm{c} = 2.0$ \cite{sl_qpt}, within an error of 1.5 percent.

\begin{figure}[!t]
\centering
\includegraphics[width=0.55\linewidth]{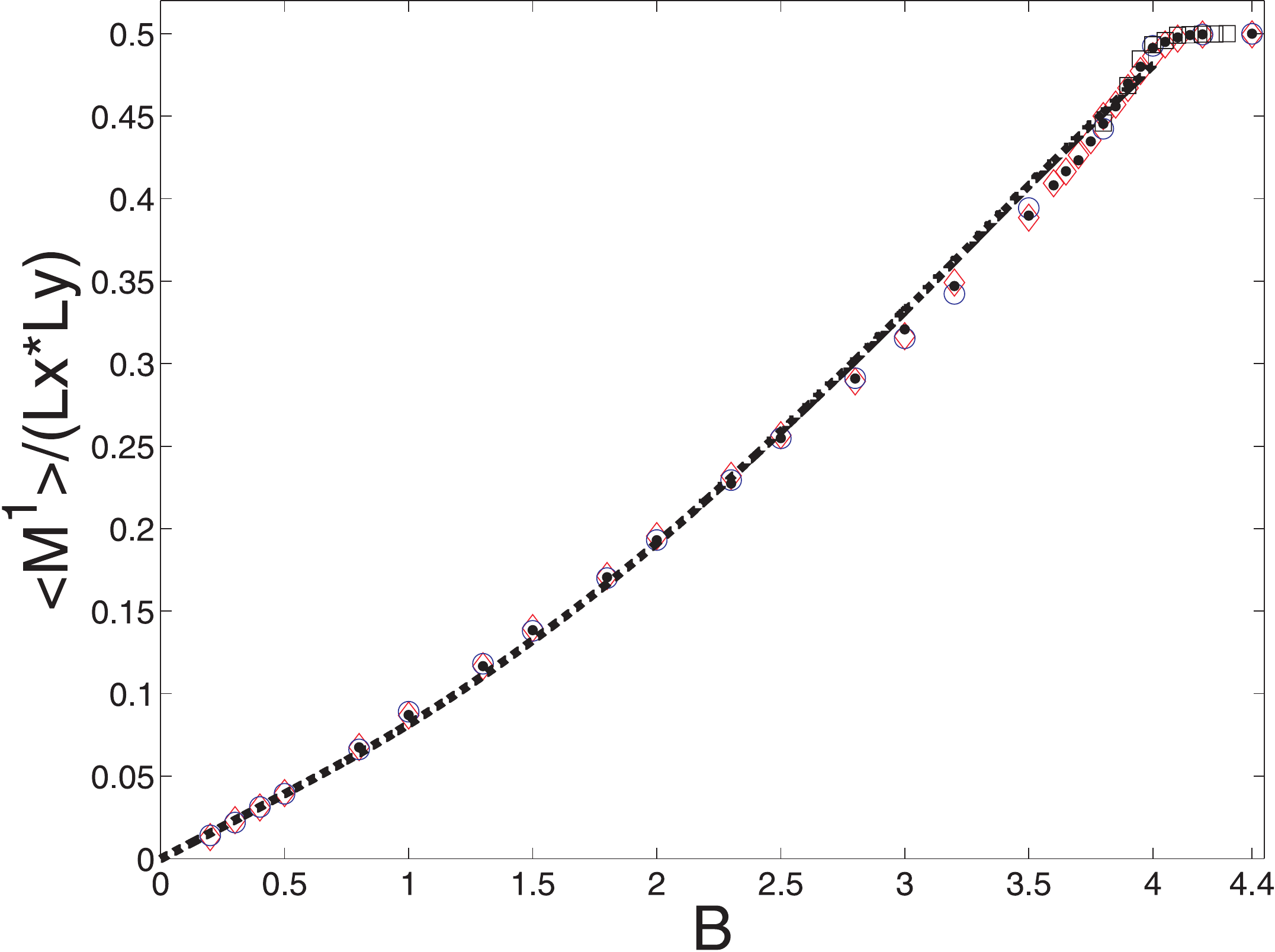}
\caption{(Color online) The magnetization density is plotted as a function of the external magnetic field $B$ for the antiferromagnetic Heisenberg model in $d=2$ dimensions very close to zero temperature for square lattices of sizes $L$. The diamond data correspond to $L = 12$ for $\beta = 15$, the empty circle and  bold dot data correspond to $L = 16$  for inverse temperature $\beta = 15$ and $\beta = 20$, respectively; and the data represented by squares correspond to $L = 24$ at $\beta = 20$. The dot line data is the one obtained using the scaling behaviour for $T = 0$ in the thermodynamic limit \cite{AFH_Scaling_behavior}.}
\label{mag2dAFH}
\end{figure}

In two dimensions, we have studied the antiferromagnetic Heisenberg model in an external magnetic field defined on spin ladders and on square lattices. In figure~\ref{mag2dAFH}, the magnetization density is displayed as a function of the external magnetic field $B$ for three different square lattices of lattice sizes $L=12, 16$ and $24$, for values of inverse temperature $\beta = 15$ and $20$. As a comparison, the numerical results obtained in \cite{AFH_Scaling_behavior} for the magnetization density in the thermodynamic limit and at zero temperature are also displayed. In this reference, a judicious law for the scaling behaviour for the ground state energy based on numerical results obtained by exact diagonalization on small lattices is proposed, and by differentiation, the magnetization in the thermodynamic limit and at zero temperature is attained. At a low external magnetic field, both results agree with each other with high accuracy, and, as $B$ increases slightly, the deviations appear, which can be conjectured due to the lack of an exact analytical expression for the scaling behaviour of the ground state stressed in that paper. For larger values of the magnetic field $B>4$, the magnetization density value saturates to its maximum value $1/2$, corresponding to a complete spin alignment. As the magnetic field goes to zero $B\rightarrow 0$, the magnetization also goes to zero, again in agreement with the Mermin-Wagner theorem. Compared to the $d=1$ case, the finite size effects are larger for large $B$-values, which is well known.

\begin{figure}[!b]
\centering
\includegraphics[width=0.98\linewidth]{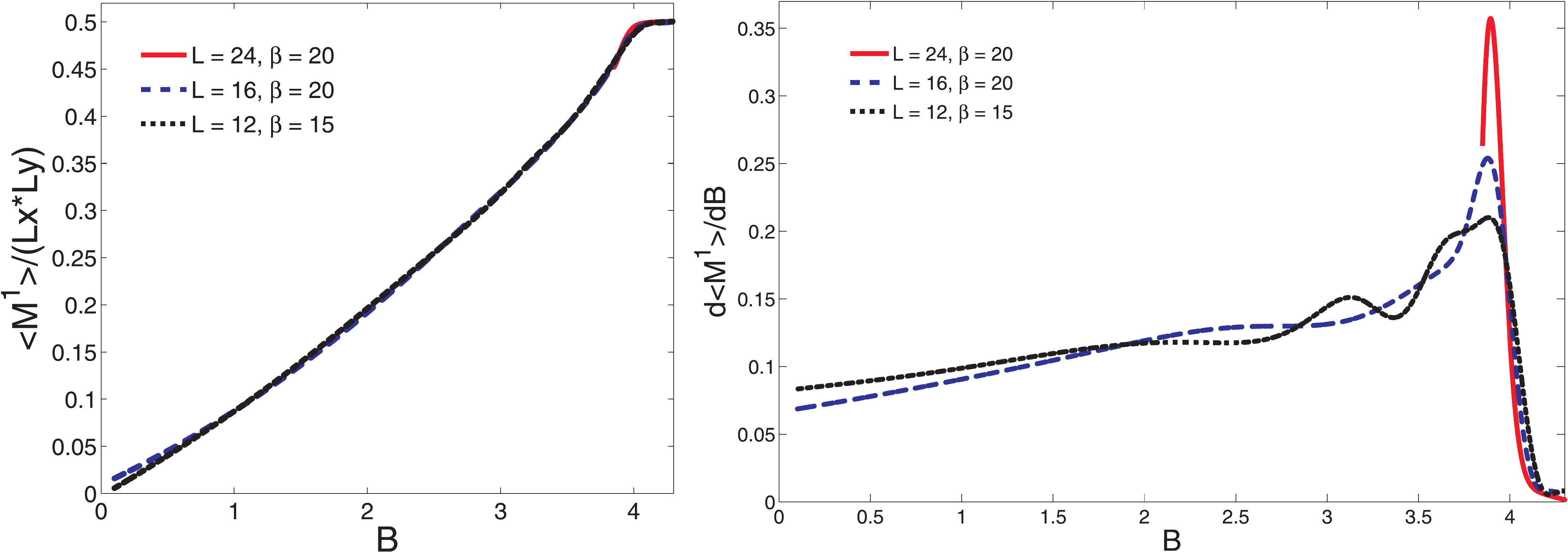}
\hspace{4cm}(a) \hspace{7cm} (b)
\caption{(Color online) (a) Fit of the numerical data plotted in figure~\ref{mag2dAFH}. (b) First derivative of the magnetization density as a function of the external magnetic field $B$ for the AFH model in two spatial dimensions.}
\label{mag_derivative2dAFH}
\end{figure}

After performing a fit to the magnetization density displayed in figure~\ref{mag2dAFH}, in a similar way as the performed in one dimension [the fit is explicitely shown in figure~\ref{mag_derivative2dAFH}~(a)], the first derivative of the magnetization density for the quantum AFH model in two spatial dimensions is attained as a function of the external magnetic field $B$, which it displayed in figure~\ref{mag_derivative2dAFH}~(b), for different system sizes and very low temperature values. It allows us to conjecture the existence of a discontinuity close to $B_\textrm{c} = 3.90$, in the thermodynamic limit and in the zero temperature limit. Within the numerical errors, this result agrees fairly well with the corresponding value reported in \cite{E. Bublitz-J. Ricardo de Sousa} for the critical value $B_\textrm{c} = 4$, with an error of 2.5 percent. Nevertheless, and in order to obtain a better accuracy, one should simulate even larger systems in two dimensions, which are unfortunately very costly by computational time.

\begin{figure}[!t]
\centering
\includegraphics[width= 0.55\linewidth]{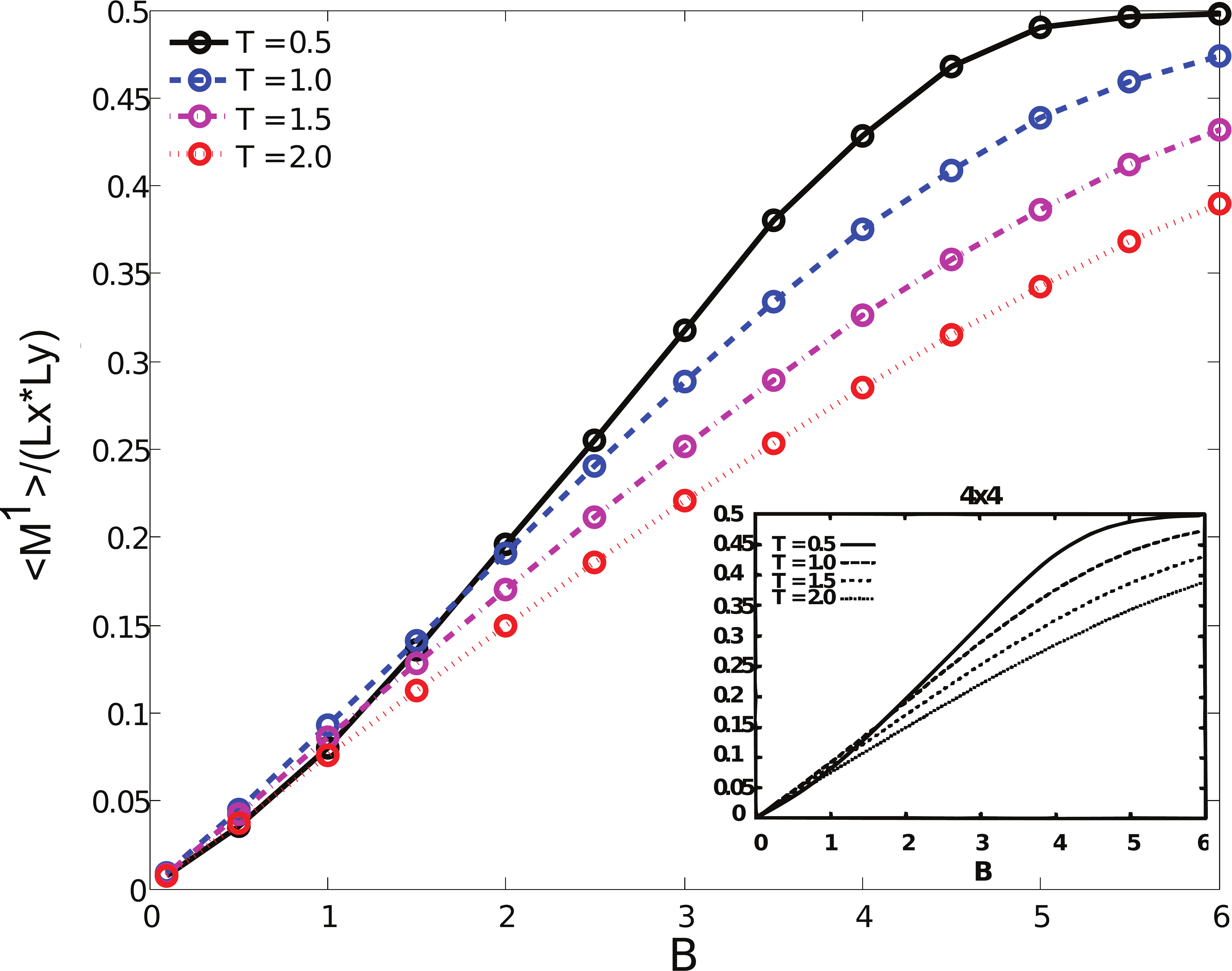}
\caption{(Color online) The high temperature behaviour for the magnetization density for the Heisenberg model in a square lattice of lattice size $L = 4$ is displayed as a function of the external magnetic field $B$ . The four curves were obtained using $\epsilon = 0.05$. The subplot corresponds to the data obtained by Fabricius et al. \cite{AFH_Scaling_behavior}.}
\label{2dMag_HighT}
\end{figure}

We also computed the magnetization of the antiferromagnetic Heisenberg model on small lattices and high temperature values. In figure~\ref{2dMag_HighT}, the magnetization density as a function of the external magnetic field is plotted for a square lattice with size $L=4$ at high temperature values $T=0.5$, 1.0, 1.5 and 2.0. As a comparison, the sub-plot shows the results from exact numerical diagonalization obtained in \cite{AFH_Scaling_behavior} for the same parameter values as the ones used in our simulations. Within a remarkable accuracy, both curves agree with each other, which shows how accurately and trustfully the meron-cluster algorithm works in the whole parameter region considered to perform numerical simulations.

Antiferromagnetic spin ladders are interesting physical systems as they interpolate between single $d=1$ spin chains and $d=2$ quantum antiferromagnets. Their low-energy dynamics are accurately described by $(1+1)-d$ quantum field theories, which can be exactly solved analytically using the Bethe Ansatz method. This correspondence was explicitly used in \cite{MCA} to compare the numerical results obtained by the meron-cluster algorithm with analytical predictions for the magnetization of spin ladder systems. We also used these results to compare with the results of our numerical simulations.

\begin{figure}[!t]
\centering
\includegraphics[width=0.55\linewidth]{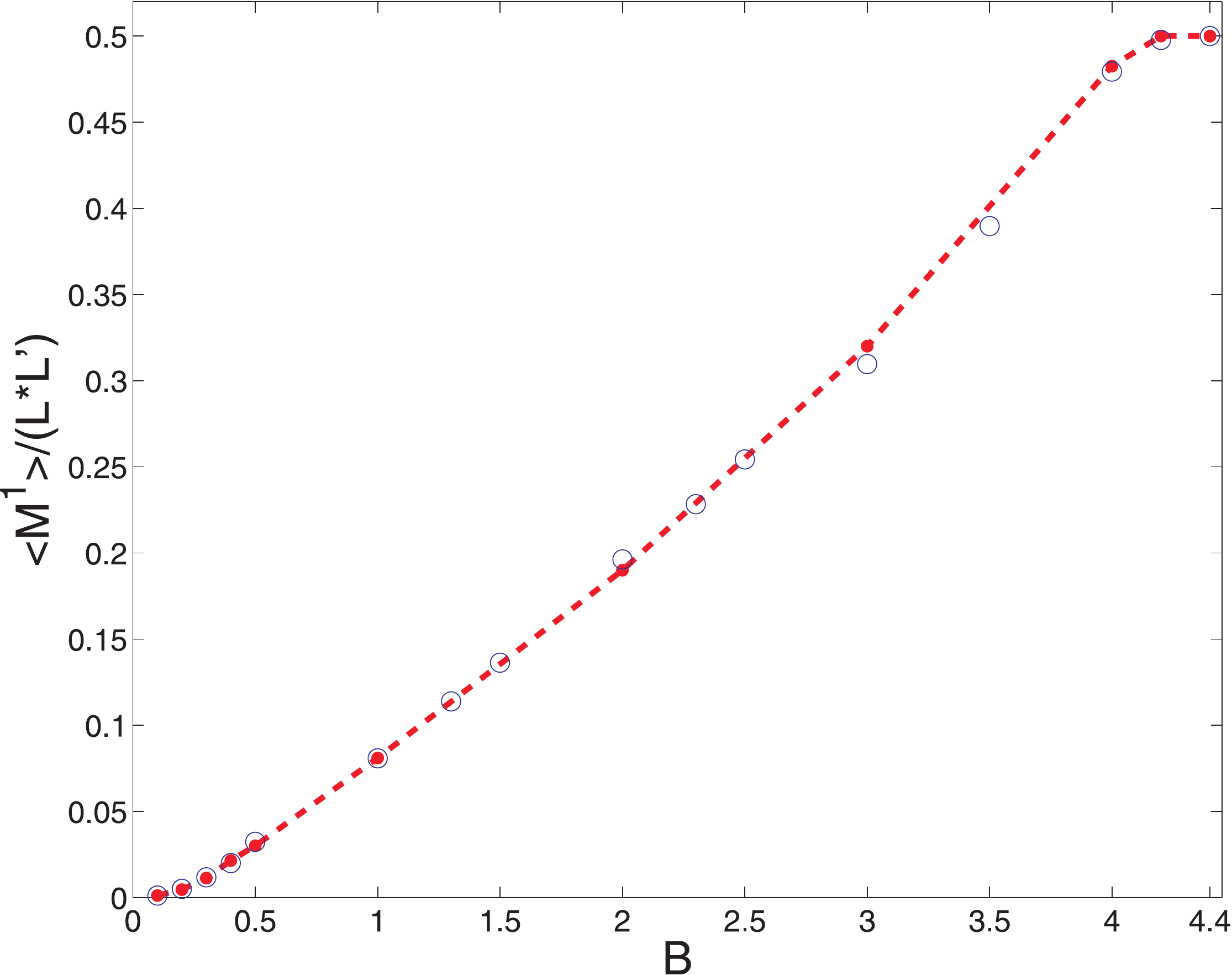}
\caption{(Color online) The magnetization density for an antiferromagnetic quantum spin ladder is plotted as a function of the external magnetic field $B$ . The circles correspond to a ladder of size $L = 20$, $L' = 4$ and inverse temperature $\beta = 15$; the bold dot with a segmented line corresponds to Chandrasekharan et al. data \cite{MCA} for the same physical system at the same $\beta$ value.}
\label{qsladder}
\end{figure}

In figure~\ref{qsladder}, the magnetization density as a function of the external magnetic field is displayed for a spin ladder of size $4\times20$ at inverse temperature $\beta = 15$. As a comparison, numerical results obtained for the same physical system by Chandrasekharan et al. \cite{MCA} are plotted. A remarkable agreement between both results is obtained in the whole range of the parameter values, which amounts to a correct implementation of the meron-cluster algorithm.

\begin{figure}[!b]
\centering
\includegraphics[width=0.55\linewidth]{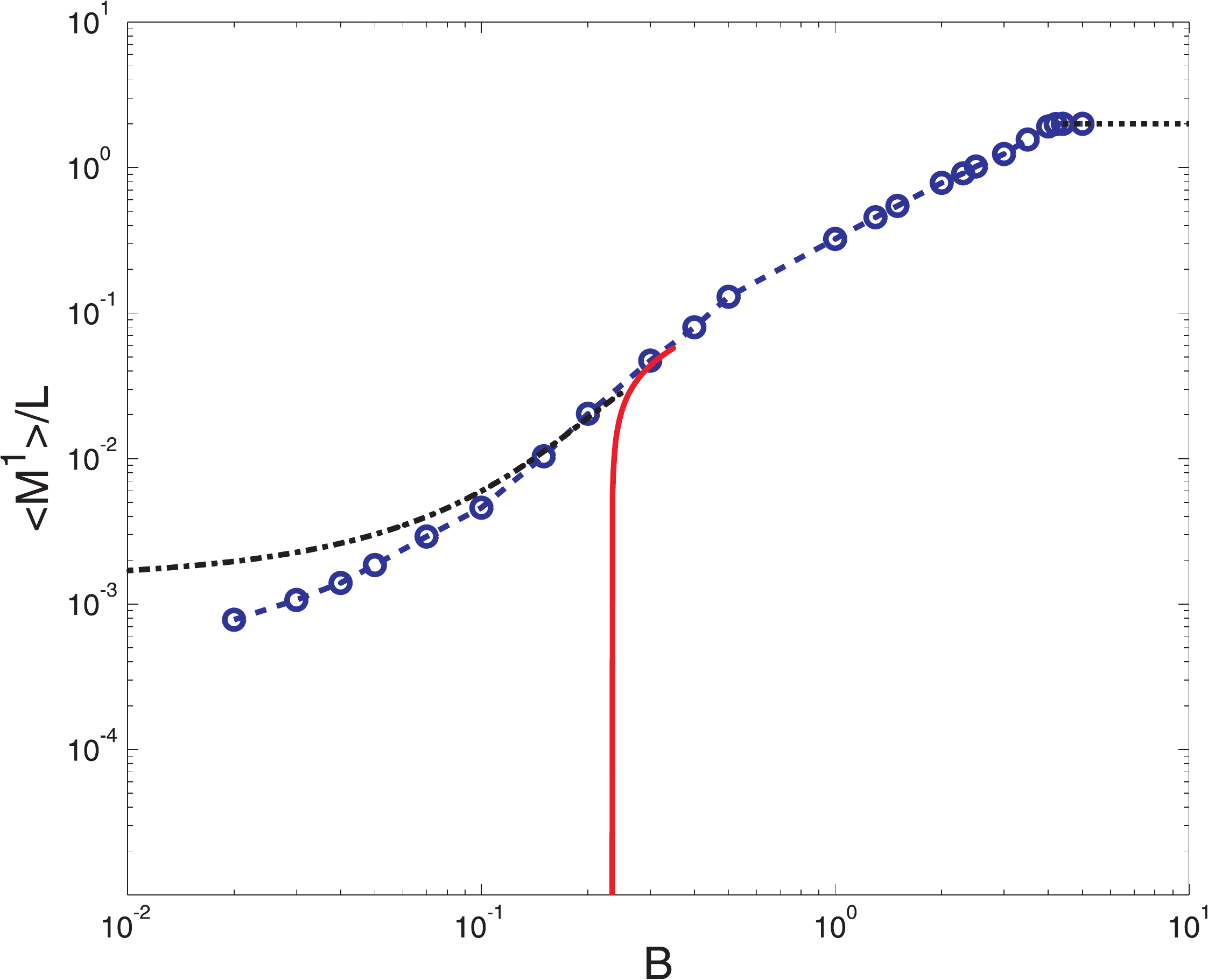}
\caption{(Color online) The magnetization density of an antiferromagnetic quantum spin ladder of dimensions  $L = 20$, $L' = 4$ and at low temperature ($\beta = 15$) is shown as a function of the external magnetic field $B$. The numerical data are represented by circle (segmented line) and it was obtained using $\epsilon = 0.075$. The segmented-dot line corresponds to the finite volume, and $T = 1/15$ analytic behaviour given by equation (\ref{analitic_lowB}) for low values of $B$, the full line curve corresponds to the infinite volume and zero temperature analytic result of equation (\ref{B_m}) for values of $B$ slightly above the threshold $ \mu \approx m$. The horizontal dotted line represents the saturation value of the magnetization density.}
\label{eff_theory_mag}
\end{figure}

The magnetization density as a function of the external magnetic field displayed in logarithmic scale is shown in figure~\ref{eff_theory_mag} for an antiferromagnetic spin ladder ($L = 20$, $L' = 4$) at low temperature $\beta = 15$. As already mentioned, spin ladders are spatially quasi-one-dimensional systems whose low-energy physics is governed by $d=(1+1)$ dimensional quantum field theories.  According to this connection, the uniform magnetic field $B$ corresponds to a chemical potential $\mu = B/c$ in the $(1+1)-d$ quantum field theory, where $c$ is the spin-wave velocity. Using the Bethe Ansatz allows one to obtain an exact solution for the magnetization in these quantum field theories \cite{Haldane}, which we will compare with numerical results along the lines proposed in  \cite{MCA}. If one constraints the simulations to an even number of coupled spin $1/2$ chains, the corresponding quantum field theory describes an asymptotically free theory with a non-perturbatively generated mass-gap $m$, and the $(1+1)-d$ $O(3)$ effective action for the spin ladder contains no topological terms.

It is known that close to the threshold region $\mu \approx m$, numerical results are very sensitive to both finite-size and finite-temperature effects. Nevertheless, in the region $\mu $, no too large a system is accurately described by a dilute fermion gas with magnetization density given by:
\begin{equation}
\frac{\langle M^1 \rangle}{L}  =\frac{1}{L} \sum_{n \in Z}\frac{1}{1+ \exp{\left\{\beta \left[c \sqrt{(2 \pi n/L)^2 + m^2}- B\right]\right\}}}\,.
\label{analitic_lowB}
\end{equation}
For chemical potential slightly above the threshold value, the magnetization density in the thermodynamic limit can be written as:
\begin{equation}
\frac{\langle M^1 \rangle}{L} = \frac{\sqrt{2m}}{\pi}\sqrt{\frac{B}{c} - m}\,.
\label{B_m}
\end{equation}
All these analytic expressions will be used to compare with our following numerical results.

The magnetization density given by equation (\ref{analitic_lowB}) is shown in figure~\ref{eff_theory_mag} as a function of the external magnetic field (dot-dashed line) in logarithmic scale. Equation (\ref{B_m}) is represented as a solid line and corresponds to its asymptotic value very close to the threshold region $ \mu \approx m$. We have used the numerical values $m = 0.141$ and $c = 1.657$ known from references \cite{m_value,c_value}, respectively. Finally, the circle-segmented line represents the numerical values obtained by the meron-cluster simulation.
It can be concluded that the numerical results agree remarkably well close to the threshold region with the behaviour predicted by the analytical expressions of the effective field theory. The deviation of the numerical results from the behaviour depicted by equation (\ref{analitic_lowB}) is due to the fact that this equation holds only for a sufficiently large number $L' \gg 1$, which is not fulfilled in the spin-ladder we have simulated. Our numerical data perfectly agree with the results obtained  in \cite{MCA}  within an error less than $2$ percent.

\begin{figure}[!b]
\centering
\includegraphics[width=0.8\linewidth]{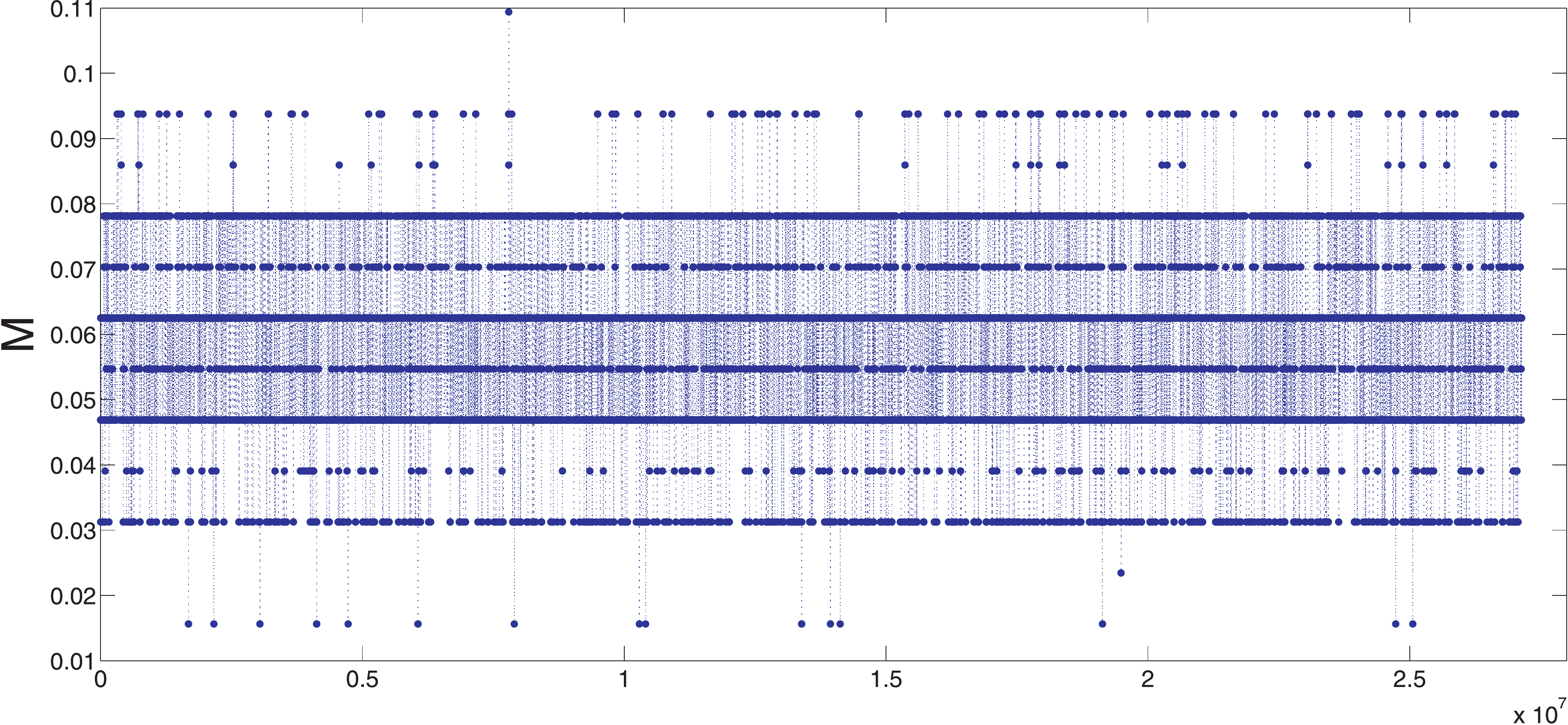}
\caption{(Color online) Time series of the magnetization density for the AFH chain of length $L = 64$ at inverse temperature $\beta = 16$ for the external magnetic field value $B = 0.5$.}
\label{time_series}
\end{figure}

The high accuracy of the numerical data obtained in the present paper, when compared with the ones obtained by other numerical methods \cite{Directed_loop,AFH_Scaling_behavior}, shows that the MCA is a reliable algorithm for Quantum Monte Carlo simulation in physical systems with dynamical competition between different types of interactions. The MCA as well as the directed loop quantum Monte Carlo represents a natural extension of the LCA and also allows us to beat the algorithm discontinuity generated in this class of systems. Nevertheless, the efficiency of the meron-cluster algorithm has not been properly measured and reported. In \cite{Directed_loop}, the authors compute the so-called integrated autocorrelation time for the magnetization $\tau_\textrm{int}(M)$ of a AFH chain of length $L = 64$, inverse temperature $\beta = 16$ and for different values of the magnetic field by using
\begin{equation}
\tau_\textrm{int}(M) = \frac{1}{2} + \sum_{t = 0}^{\infty} \Gamma_M(t),
\label{int_time_exp}
\end{equation}
where $\Gamma_M(t)$ is the autocorrelation function of the magnetization. Typically,
the autocorrelation function is not a simple exponential function of time. This fact makes it rather difficult to estimate the integrated autocorrelation time given by equation (\ref{int_time_exp}). Another method to estimate the integrated autocorrelation time of an algorithm is to examine how many sweeps in the stochastic process are necessary to get statistically independent measurements for an observable, which can be achieved by performing a binning method \cite{binning,binning_2}. In order to measure the efficiency of the MCA,  we have applied a binning method to the time series of the magnetization for the AFH chain using the same values (size and temperature) as the ones used in reference \cite{Directed_loop}, and for an external magnetic field of the value $B = 0.5$. The length of the stochastic time series generated by the Markov chain was $ n = 2.75 \times 10^7$. We denote this array by $M$, and its shape is shown in figure~\ref{time_series}.

\begin{figure}[!t]
\centering
\includegraphics[width=0.8\linewidth]{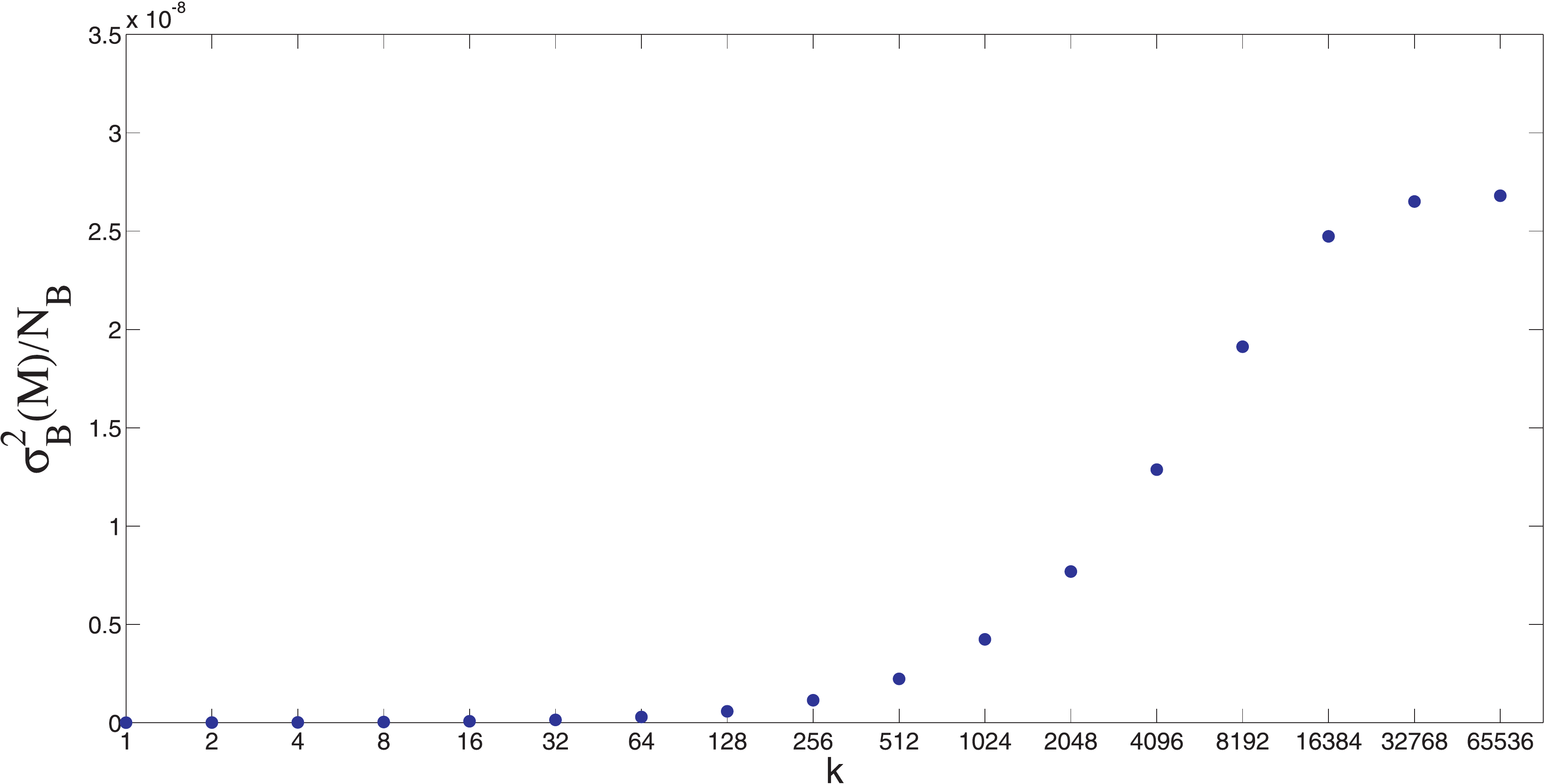}
\caption{Variance over the number of blocks as a function of the bin size $k$ for the magnetization density of the AFH ring of length $L = 64$ at inverse temperature $\beta = 16$ for an external magnetic field value $B = 0.5$. The convergence criterion is achieved up to $k = 2^{15}$, which ensures that $\bar{M}_{{B}i}$ becomes statistically independent among the $N_{B}$ blocks}
\label{binning_criterion}
\end{figure}

Next we divide the time series into blocks, each one of length $k$, where $k = 2^1, 2^2, 2^3, \ldots$, the number of blocks $N_{B}$ is a function of the binning size $k$ and is given by $N_{B} = [n/k]$, where $[\hspace{0.2cm}]$ denote the integer part. Any remaining values of the time series after this binning process are disregarded.
The variance $\sigma_{B}^2(M)$ of the block average on each block $ \bar{M}_{{B}i}$ with respect to the average of all blocks together $\bar{M}_{B}$ is given by:
\begin{equation}
\sigma_{B}^2(M) = \frac{1}{N_{B} -1} \sum_{i=1}^{N_{B}} \left( \bar{M}_{{B}i} - \bar{M}_{B} \right)^2,
\end{equation}
which is a function of the bin size $k$. This variance should become inversely proportional to $k$ as the bin size $k$ becomes large enough, so that $\bar{M}_{{B}i}$ becomes statistically independent among the blocks. This can be numerically checked by plotting $\sigma_B^2/N_{B}$ over several bin sizes. This quantity allows us to estimate the error and is shown in figure~\ref{binning_criterion}. In this figure, one can see that for values of bin sizes over $2^{15}$, the statistical independent regime is already reached. Moreover, the error relative to this quantity between bin sizes $2^{15}$ and $2^{16}$ is of the order of $1.1\%$, which shows that the averages of each block are statistically independent of each other.

The main aim of the binning method is to analyse how many bin sizes $k$ are necessary to get statistically independent values of the average of each block $\bar{M}_{{B}i}$. This can be translated back into the Monte Carlo method. In this method, an observable is measured by using the importance sampling generating a Markov process, which fulfils a detailed balance and ergodicity. The expectation value of an observable $\langle \hat{O} \rangle$ is estimated by the simple arithmetic average of the observable evaluated in these configurations $\bar{O}$. If we repeat the whole Monte Carlo simulation many times, we can get an estimator of the observable in each of these Monte Carlo runs $\bar{O}_i$. If the data are uncorrelated, then, by the central limit theorem, the distribution of $\bar{O}_i$ is Gaussian related to its mean value, and, therefore, the estimator of the observable is a reliable estimator of the expectation value $\langle \hat{O} \rangle$:
\begin{equation}
\bar{O} \approx \langle \hat{O} \rangle.
\end{equation}
The numbers of Monte Carlo runs generated are equivalent to the numbers of block $N_{B}$, and similarly the number of sweeps on each
single Monte Carlo simulation is equivalent to the bin size $k$ of the binning method. As for binning sizes up to $k = 2^{15}$, the
averages of blocks become statistically independent. Therefore, the arithmetic mean over the Markov chain can reliably estimate the expectation value of the observable $\langle \hat{O} \rangle$ up to the number of sweeps of the order of $100\times 2^{15}$.

\begin{figure}[!t]
\centering
\includegraphics[width=0.8\linewidth]{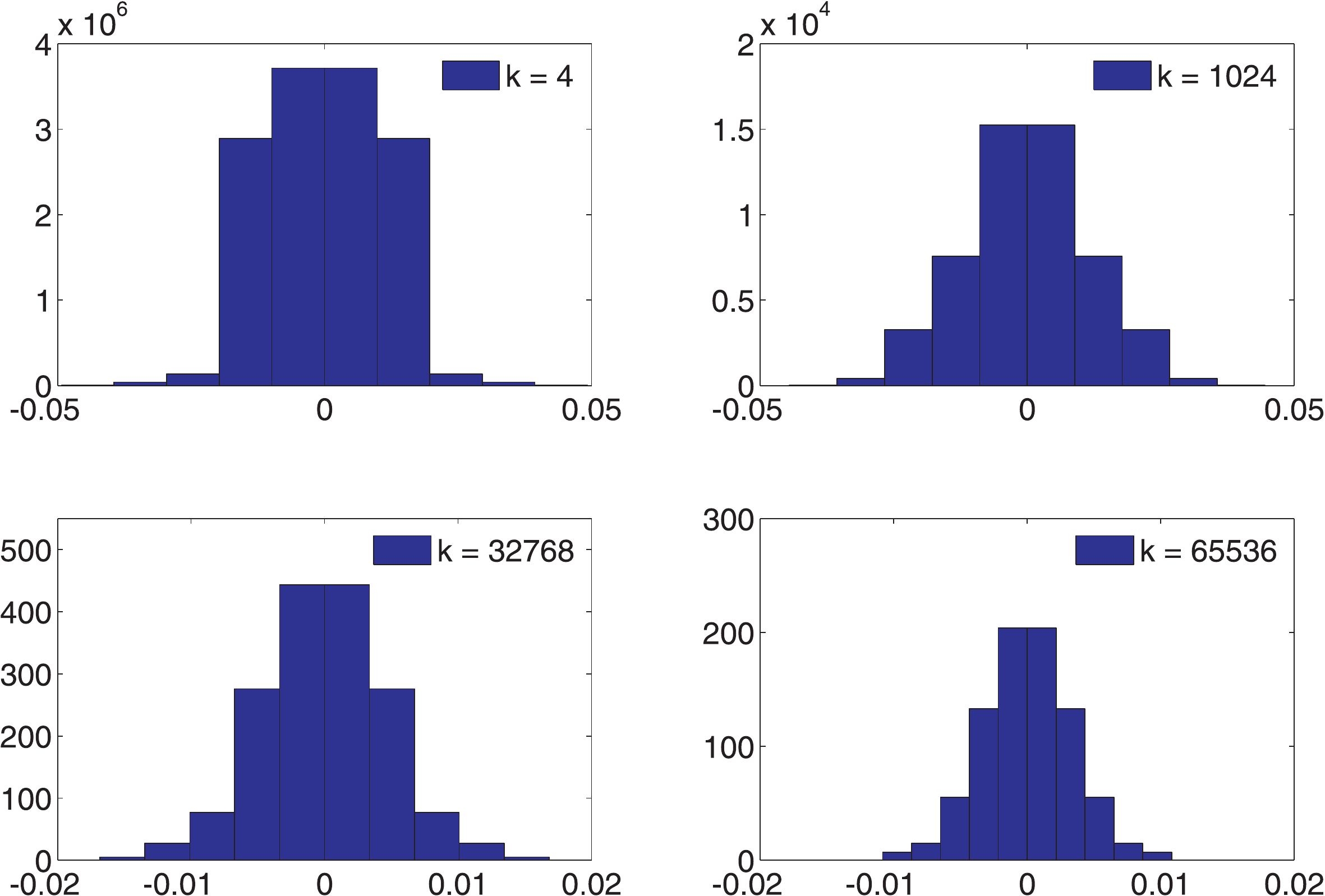}
\caption{(Color online) Probability distribution of the averages of blocks related to their mean value for four different bin sizes $k$.}
\label{Sim_block_distri}
\end{figure}

In figure~\ref{Sim_block_distri}, we show the probability distribution of the averages of blocks related to their mean value for different values of bin sizes $k$. One can see that from bin size up to $k = 2^{15}$ statistically independent data are generated, and the distribution approaches the regime of the central limit theorem. Finally, we estimate the integrated autocorrelation time of the MCA for the magnetization $\tau_\textrm{int}(M)$ using the binning analysis. It holds that \cite{binning,binning_2}:
\begin{equation}
\sigma^2_{\bar{O}_i} = \frac{\sigma^2_{O_j}}{r}2\tau_\textrm{int}(O),
\end{equation}
where $\sigma^2_{\bar{O}_i}$ is the variance of the estimator obtained in each  Monte Carlo run, and $\sigma^2_{O_j}$ is the variance of the individual measurement. We consider that in a single Monte Carlo simulation, $r$ sweeps are generated. Hence, from figure~\ref{binning_criterion}, the integrated autocorrelation time can be estimated by:
\begin{equation}
2\tau_\textrm{int}(O) = \frac{\left\lbrace \sigma_B^2(O)/N_{B} \right\rbrace_{k \rightarrow \infty}}{\left\lbrace \sigma_B^2(O)/N_{B} \right\rbrace_{k = 1}}
\label{tau_binning}
\end{equation}
and, therefore, using equation (\ref{tau_binning}) and from figure \ref{binning_criterion}, the estimation of the integrated autocorrelation time of the MCA for the magnetization of the AFH ring of length $L = 64$ at the inverse temperature $\beta = 16$ for an external magnetic field value $B = 0.5$ is:
\begin{equation}
\tau_\textrm{int}(M) = 2889.500 \pm 0.013 ,
\end{equation}
where the error was computed by using the expression $\sigma_{B}(M)/\sqrt{N_{B}}$ for $k \rightarrow \infty$. The value obtained for the integrated autocorrelation time is very large and in practice would limit the efficiency of the MCA for large systems. By comparing this value with the one obtained in \cite{Directed_loop}, one can see that the directed loop method is more efficient than the MCA. Nevertheless, both algorithms give very accurate numerical results for the simulated physical systems. The relative large value for $\tau_\textrm{int}(M)$ is related to the fact that only configurations from the zero-meron sector, which is much smaller than the meron sector, do contribute to the magnetization, which in practice represents a strong constraint in the process of building clusters leading to a relative low acceptance rate of changes of new strings within a cluster. This limitation partially spoils the efficiency of the algorithm.

\section{Conclusions}

In this article, we have numerically simulated the antiferromagnetic quantum Heisenberg model in a uniform external magnetic field in one and two spatial dimensions by using the meron-cluster algorithm proposed in \cite{MCA}. Our results agree remarkably well with the previous results obtained in \cite{AFH_Scaling_behavior} by using suitable scaling behaviour formulae for the energy eigenvalues and performing numerically a thermodynamic limit. We have also simulated the quantum spin ladders in a magnetic field studied in \cite{MCA}, and have found a perfect agreement with their numerical results with numerical differences less than a few percent. The present study allows us to confirm that the meron-cluster concept applied to the quantum spin systems in the presence of an arbitrary external magnetic field represents a trustable method to simulate these kinds of systems, even when the independence cluster-flip feature, which is achieved when the quantization is performed in the perpendicular direction to the magnetic field, leads to a sign problem. Since the sign problem also appears in bosonic and fermionic systems in higher dimensions than one, the present application could shed some light to the numerical study of those systems. Moreover, its efficiency encourages us to apply it for the study of systems presenting a dynamical competition between different interactions, which can produce in the thermodynamic limit a non-unique ground state at a very low temperature, which is a characteristic feature of quantum phase transitions.

As the limits of infinite volume and zero temperature are reached, the magnetization density collapses into an asymptotic curve, and the peak of its corresponding first order derivative allows us to suggest that there is a quantum phase transition close to the critical value $B_\textrm{c} = 1.97$ and $B_\textrm{c} = 3.90$ in one and two spatial dimensions, respectively. These results are in fairly good agreement with the results obtained by the exact diagonalization method in one dimension \cite{sl_qpt}, and the mean field theory in three spatial dimensions \cite{E. Bublitz-J. Ricardo de Sousa}, which corresponds to the AF $(2+1)$-dimensional quantum Heisenberg model we have simulated.
Nevertheless,  in order to address these phenomena with even more precision, new and more accurate simulations should be performed, which we plan to carry out in future investigations.

In practice, some limitations arise when studying more carefully the efficiency of the meron-cluster algorithm. Although the MCA gives very accurate numerical results for the models considered in this paper,  which are in a remarkable agreement with the ones obtained by other methods, in view of the estimated value for the integrated autocorrelation time, one can conclude that the MCA is not as efficient as the directed loop quantum Monte Carlo method or the worm algorithm for the study of these classes of physical systems. It will be a very interesting question whether the directed loop approach could be used to further extend the applicability of the meron concept for solving sign problems in other models \cite{Directed_loop}.

\vspace{-3mm}

\section*{Acknowledgements}

This work was partially supported by Dicyt-USACH Grant No.~041531PA. A.R. acknowledges Conicyt Grant No.~21090138 for financial support during his Ph.D. study. The authors would like to thank \mbox{U.-J.~Wiese} for enlightening physical discussions concerning the meron-cluster algorithm and D.~Zambrano for valuable hints concerning programming strategies, and H.G.~Evertz for useful discussions concerning the binning method. G.P. would like to thank the Institute for Theoretical Physics at Bern University and their members for their kind hospitality. We also acknowledge the University of Bern Switzerland for allowing us to use its cluster of computers.

\newpage

\ukrainianpart

\title{Мерон-кластерне моделювання квантової антиферомагнітної моделі Гайзенберга у магнітному полі в одному і двох вимірах}
\author{Г. Палма, А. Ріверос}
\address{Відділ фізики, Універсідад де Сантьяго-де-Чилі, Сантьяго 2, Чилі}

\makeukrtitle

\begin{abstract}
Будучи вмотивованими числовими симуляціями систем, які проявляють властивості квантових фазових переходів, представляємо нове застосування
мерон-кластерного алгоритма з метою моделювання квантової антиферомагнітної моделі Гайзенберга, що взаємодіє із зовнішнім однорідним магнітним полем, як в одно- так і в двовимірному випадках. В границі нескінченного об'єму і при нульовій температурі нами знайдено числові докази, які підтверджують наявність квантового фазового переходу дуже близько до критичних значень $B_\textrm{c}=2$ і $B_\textrm{c} = 4$ для системи, відповідно, в одно- і двовимірному випадках. Для одновимірної системи нами здійснено порівняння числових даних, отриманих з аналітичних результатів для густини намагнічування як функції зовнішнього поля, отриманої за допомогою аналізу скейлінгової поведінки та техніки анзацу Бете.
Оскільки не має аналітичного розв'язку для двовимірного випадку, автори порівняли свої результати з густиною намагнічування, отриманою за допомогою скейлінгових співвідношень для невеликих розмірів граток і з апроксимованою термодинамічною границею при нульовій температурі, оціненою зі скенйлінгових співвідношень. Більше того, автори порівняли числові дані з іншими числовими симуляціями, виконаними з використанням різних алгоритмів в одному та у двох вимірах, типу методу напрямлених петель. Отримані числові дані ідеально узгоджуються з усіма попередніми рузультатами, а це в свою чергу підтверджує, що мерон-алгоритм є надійним для проведення моделювання методом Монте Карло і його можна застосовувати в одному і двох вимірах. І нарешті, нами обчислено сумарний автокореляціний час з метою вимірювання ефективності мерон-алгоритма в одному вимірі.
\keywords симуляції кластерним алгоритмом, проблема знаку, квантові фазові переходи, методи квантового Монте Карло
\end{abstract}

\end{document}